\documentclass[useAMS,usegraphicx,usenatbib]{mn2e}

\bibpunct{(}{)}{;}{a}{}{,}

\def\rcm2{\rm cm^{-2}}

\def\chiq{$\chi^2$}
\def\sax{\mbox{\emph {BeppoSAX}}}
\def\xte{\emph{RXTE}}
\def\j1859{\mbox{XTE~J1859+226}}
\def\1550{\mbox{XTE~J550--564}}

\def\ktbb{kT_{\rm bb}}

\def\ktin{kT_{\rm in}}

\def\kte{kT_{\rm e}}
\def\rin{R_{\rm in}}
\def\nh{N_{\rm H}}

\def\ef{E_{\rm f}}
\def\nbmc{N_{\rm bmc}}

\def\wabs{{\sc wabs}}
\def\comptb{{\sc comptb}}
\def\compps{{\sc compps}}
\def\comptt{{\sc comptt}}
\def\eqpair{{\sc eqpair}}
\def\bmc{{\sc bmc}}
\def\bb{{\sc bb}}
\def\dbb{{\sc diskbb}}

\title[Spectral evolution of \j1859]{Spectral evolution of the X-ray nova XTE J1859+226 during its outburst  observed
by \emph{BeppoSAX} and \emph{RXTE}}
\author[R. Farinelli et al.]{R. Farinelli$^{1,2}$, L.~Amati$^{3}$, N.~Shaposhnikov$^{4}$, F. Frontera$^{1}$, N. Masetti$^{3}$, E. Palazzi$^{2}$, 
 \newauthor  R.~Landi$^{3}$, C.~Lombardi$^{1}$, M.~Orlandini$^{3}$, and C. Brocksopp $^{5}$
\\
\\
$^{1}$Dipartimento di Fisica, Universit\`a di Ferrara, via Saragat 1, 44122 Ferrara, Italy\\
$^{2}$INAF-IASF, Sezione di Palermo, via U. La Malfa 153, 90146 Palermo, Italy\\
$^{3}$INAF-IASF, Sezione di Bologna, via Gobetti 101, 40129 Bologna, Italy\\
$^{4}$NASA-GSFC, Astrophysics Science Division, Greenbelt, MD 20771, USA\\
$^{5}$Mullard Space Science Laboratory, University College London, Holmbury St Mary, Dorking, Surrey RH5 6NT\\
}

\begin{document}

\date{Submitted .}

\pagerange{\pageref{firstpage}--\pageref{lastpage}} \pubyear{2008}

\maketitle
        
\label{firstpage}

\begin{abstract}
We report  results of an extensive analysis of the \mbox{X-ray} nova \j1859\ observed with \sax\ and \xte\ 
during its 1999 outburst. 
We modelled the source spectrum with a multicolour blackbody-like feature  plus the generic Comptonization model \bmc\  which has the advantage of providing 
spectral description of the emitted--radiation properties without assumptions on the underlying
physical process. The multicolour component is attributed to the geometrically thin accretion disk, while the Comptonization 
spectrum is claimed to originate in the innermost sub-Keplerian region of the system (transition layer).
We find that \j1859 covers all the spectral states typical of black-hole sources during its 
evolution across the outburst, however during the very high state, when the disk contribution to the total luminosity is more than 70\%
and the root mean square variability $\la$ 5\%, the high-energy photon  index is closer to a hard state value ($\Gamma \sim 1.8$).
The \bmc\ normalization and photon index $\Gamma$ well correlate with the radio emission,
and we also observed a possible saturation effect of $\Gamma$ at the brightest
radio emission levels.
A  strong positive correlation was found between the fraction of Comptonized seed photons 
and the observed integrated root mean square variability, which  strengthens the idea that 
most of the fast variability in  these systems is provided by the innermost Compton cloud,
 which may be also identified as a jet.

\end{abstract}


\section{Introduction}
\label{introduction}

X-ray novae, or soft X-ray transients (SXTs), are low--mass \mbox{X-ray}
binaries
which undergo a sudden and 
unpredictable 
increase in intensity up to 2--6 orders of
magnitude in \mbox{X-rays} \citep{liu01}. These outbursts
occur  usually on time--scale of decades, even if there are cases like GX 339--4 where 
regular annual outbursts are seen \citep{mcr06}.
Their light curves are characterized by a typical rise--time 
of few days and an e--folding decay time of several tens of days.
In these systems the accreting compact
object is found to be a neutron star (NS) or a black hole (BH)
 \citep[e.g.,][]{ts96,Cherepashchuk00}.
During the outbursts,
thought to be originated by a dramatic change in the mass
accretion rate, the \mbox{X-ray} spectrum evolves through different states,
classified according to the presence and relative prominence of a soft thermal 
component
and a hard tail component \citep[e.g.,][]{zg04,rm06, cadolle06, delsanto08, laurent11}.
The thermal component is commonly thought to be generated
in the inner regions of the accretion disk, and the non thermal
hard tail by Comptonization of the soft thermal photons by a hot 
electron corona \citep[e.g.,][]{ps96}.
The \mbox{X-ray} spectra of several SXTs may  allegedly show  a reflection component and/or 
emission lines and edges, 
due to reprocessing of the hard \mbox{X-ray} 
photons by the cold outer regions of the accretion disk \citep[e.g.][]{Magdziarz95}.
The power spectra derived from the \mbox{X-ray} light curve
of SXTs in outburst are characterized by a red noise component with
superimposed quasi-periodic oscillations (QPOs) at different and quickly varying frequencies.
Black hole SXTs show generally no power above $\sim$100 Hz and low frequency
QPOs, while the power spectra of NS SXTs extend to higher 
frequencies and
typically show kHz QPOs \citep[e.g.,][]{Vanderklis00,Belloni02}.

The \mbox{X-ray} outburst is accompanied by an optical outburst, with
typical ratio between \mbox{X-ray} and optical luminosities of $\sim$1000.
The reprocessing of the \mbox{X-ray} emission from the outer regions of the disk 
is a possible way to explain the UV, optical and IR emission \citep[e.g.,][]{Vanparadijs95,Esin00, coriat09, gandhi11}. 
By means of optical spectroscopy in quiescence, it has been
possible to determine the mass function of several of these systems, which
is the most direct way to establish if the compact accreting object is a
 NS or BH. 
Radio emission has been detected from several SXTs in outburst, and in some
systems, often referred to as "microquasars", jets could be resolved \citep{fender06}.
Simultaneous \mbox{X-ray} and radio observations of SXTs, albeit the global picture
is not yet completely clear, seem to show some properties shared
by almost all sources. 
In the canonical \mbox{X-ray} hard state, a steady and powerful jet is
always observed with moderate velocity ($\Gamma \leq 2$), an associated flat radio spectrum ($\alpha \sim 0$)
extending beyond the radio band and linear polarization
level of a few percent. During the source transition from hard to soft state,
major radio flares are observed with ejection events stronger 
($\Gamma \geq 2$) than the hard state. The radio emission gets subsequently
quenched as the source has reached the soft state, and newly turns-on
in the soft to hard transition at later stages of the outburst \citep{fender09}.
Albeit classical studies of jets in SFXTs have been performed in the radio band,
in the recent years it has been shown that  the contribution to the optical and infrared emission
can be carried-out for a significant fraction by the jet, both in black hole
and neutron star \mbox{X-ray} binaries \citep{russell06}.
Strong indirect evidence for optical and infrared fast-variable emission from
a jet has been reported by \cite{malzac04} and \cite{hynes06}.

In this paper we report and discuss the results of an observational
campaign of \j1859\ performed with \sax\ and \xte\ during the source 1999
outburst.
In Section \ref{literature} we give an overview of the source properties
from the literature, in Section \ref{sax_xte_analysis} we report
information on the periods of observations with the two satellites and on the data reduction procedure, 
in Section \ref{results} we show the main results of the spectral
analysis, in Section \ref{discussion} we discuss the implications
for the source accretion geometry, some considerations on the relation
between the high-energy spectral features and the radio emission,
 the main observational difference with the systems hosting a neutron star and the origin of the
fast temporal variability. Finally we draw our conclusions in Section \ref{conclusions}.

\begin{table*}
\label{log_table}
\caption{Log of \sax\ TOO observations of \j1859. For each NFI, the left column reports the time exposure
(ks), the right column the mean count rate (counts~$s^{-1}$) in the given energy band.}
\smallskip
\small
\begin{center}
\begin{tabular}{l l l|c c c c|c c c c }
\hline
\hline
TOO &  Epoch  & MJD  &  \multicolumn{2}{c|}{LECS} & \multicolumn{2}{c}{MECS} & \multicolumn{2}{c|}{HP} & \multicolumn{2}{c|}{PDS} \\
 &     &     &  \multicolumn{2}{c|}{0.15-4 keV} & \multicolumn{2}{c}{1.5-10 keV} & \multicolumn{2}{c|}{8-30 keV} & \multicolumn{2}{c|}{15-200 keV} \\

\hline
1 & Oct. 15, 1999  & 51466 & off &  -- & 14.1 &  $181.1\pm0.1$ & 14.7 &  $63.6\pm0.2$  &  7.1 &  $42.6\pm0.1$  \\ 

2 &Oct. 22, 1999 & 51473 & off &  --  &   16.1 &  $159.1\pm0.1$ & 17.2  & $48.8\pm0.1$  & 7.7  & $37.9\pm0.1$ \\    
3 &Oct. 28, 1999 & 51479 & 0.4 & $163.4\pm 0.7$   & 4.4   &  $223.6\pm0.2$ & 4.0  & $45.8\pm0.3$  & 1.8  & $29.9\pm0.2$ \\    
4 &Nov. 7, 1999 &  51490  & 0.7 & $139.5\pm 0.5$    &  28.8  &  $135.9\pm0.1$ &  31.4 &  $9.6\pm0.1$ &  13.6 &  $8.2\pm0.1$\\    
5 &Nov. 19, 1999 & 51501 & 4.5 & $73.9\pm 0.1$   &  30.1  & $81.3\pm0.1$  &  32.3 &  $4.3\pm0.1$ &  12.8 & $8.2\pm0.1$ \\    

6 &Mar. 25, 2000 & 51628 & 18.2 & $0.80\pm 0.01$   &  40.8  & $0.97\pm0.01$  & 45.2  &  $0.97\pm0.10$ & 21.3  &  $1.14\pm0.01$\\    
\hline
\hline
\end{tabular}
\end{center}
\end{table*}

\section{The outburst of \j1859}
\label{literature}

\begin{figure}
\centerline{\includegraphics[width=6.5cm, height=9cm, angle=-90]{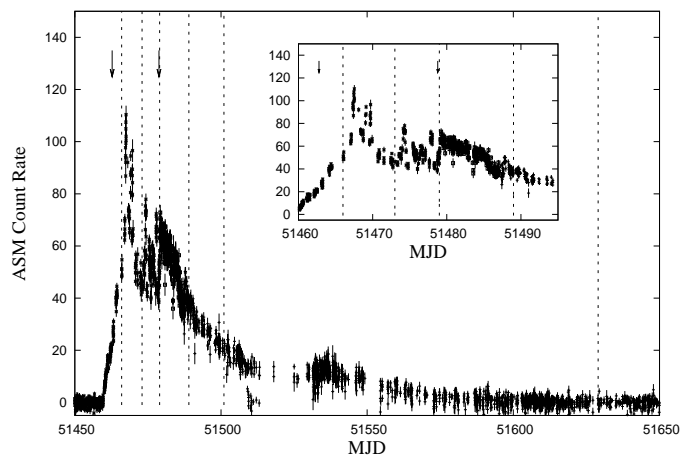}}
\caption{\emph{RXTE/ASM} light curve of the entire \j1859\ outburst and zoom
of the of the interval MJD 51460-51495 showing the flaring activity of the source before the beginning
of the e-folding decay phase.
The dashed vertical lines indicate the epochs of the \sax\ observations, while the two arrows show
the interval of data analysed by ST09.}
 \label{asm_lc}
\end{figure}

The transient source  \j1859\ was first detected by the All--Sky--Monitor (ASM)
aboard of the \emph{Rossi X-Ray
Timing Explorer} (\xte) satellite on 1999 October 9
(MJD 51460) at the Galactic coordinates 
$l=54.05^{\circ}$, $b=+8.61^{\circ}$ 
with an intensity
of $160\pm15$ mCrab in the 2--12 keV band \citep{Wood99}.
The \emph{ASM} light curve (Fig. \ref{asm_lc}) shows a structure reminiscent
of a Fast--Rise and Exponential--Decay (FRED) shape typical of SXTs, 
with a rise--time (from $10\%$ to $90\%$ of the peak) of about 5 days, a maximum intensity of 1.4 Crab and
a decay--time (\emph{e--fold}) of $\sim23$ days.
This behaviour is however not smooth, and in the temporal interval from MJD 51468 to
MJD 51479 during the decay phase, a strong flaring activity occurred.
 
The outburst of \j1859\ was also revealed by the 
\emph{BATSE} experiment  onboard the \emph{Compton Gamma Ray Observatory}  
up to 200 keV, with a peculiar behaviour: the hard \mbox{X-ray} flux reached
its maximum (MJD 51462) while the soft \mbox{X-ray} flux (ASM) was still increasing
\citep{mcw99}.
Analysing the soft and hard \mbox{X-ray} temporal behaviour and comparing it with
other SXTs, \citet[][hereafter B02]{brocksopp02}, concluded that the hard state 
observed in this source at the outburst rise  (despite the name of ``soft'' \mbox{X-ray} transients), 
rather then being a peculiar behaviour of \j1859, can be an ubiquitous feature for this class of objects.

The temporal properties of the source were thoroughly 
investigated by \citet[][hereafter C04]{casella04} using extensive \xte\ observations 
(from MJD 51462 to MJD 51546). The Power Density Spectra (PSD) obtained
with the Proportional Counter Array (PCA, 3-50 keV) showed
three main types of low-frequency (1-10 Hz) QPOs, 
according to the centroid frequency value, the coherence parameter Q, 
the QPO amplitude, and the shape of the underlying
noise (see Table 1 in C04).
For C-type QPOs (the ones having $Q \ga 10$), a strong anti-correlation was found 
between the frequency and the root mean square (RMS) variability, while the RMS amplitude of
all the three classes increases with energy, flattening above 10 keV, ruling
out a disk origin.

The source was also detected 
during its quiescent state in February 5, 2003, by the Chandra observatory;
the 0.75--2.7 keV spectrum could be fitted by a simple power--law with
a poorly constrained photon index $\Gamma \sim$2.4  \citep{Tomsick03}. 
The optical counterpart of \j1859\ was discovered 
on 1999 October 12 at 
the position of R.A.(2000)~=~$18^{\rm h}58^{\rm m}41^{\rm s}.58$, 
Dec(2000)~=~+$22^{\circ}39^{\prime}29^{\prime \prime}.4$, with an 
$R$--band 
magnitude of $\sim$ 15.1  \citep{Garnavich99} and was
observed with different optical and infrared telescopes, including
HST and UKIRT \citep{Hynes02}. 
The optical spectrum
was characterized by a strong blue continuum with weak emission lines 
typical of an \mbox{X-ray} nova in outburst \citep{Garnavich99,Wagner99,Sanchezfernandez01,Zurita02}.
Optical light curves present a FRED behaviour  similar to that observed 
in the soft 
\mbox{X-ray} band, with maximum 
magnitude $V\sim15$ and quiescent magnitude $V\sim23$, a decay
rate of $0.018\pm0.002$ mag per day. Evidences of 3 mini--outbursts 
near quiescence \citep{Sanchezfernandez01,Zurita02}, a reflare at
about 50 day after the maximum and possible superhamps \citep{Uemura04},
were also found. 

A strong radio counterpart of the source was discovered on 1999 
October 11 at a flux density of $\sim$ 10 mJy at 15 GHz with the Ryle Telescope and 
\emph{VLA} \citep{Pooley99}. The source position  
of R.A.(2000)~=~$18^{\rm h}58^{\rm m}42^{\rm s}.0$,
Dec(2000)~=~+$22^{\circ}39^{\prime}10^{\prime \prime}.0$ is
consistent with the position of the optical counterpart.
The radio light curve of \j1859\ was characterized by five major
radio flares superposed to a somewhat global
fading behaviour (B02). To first approximation, the sequence
of radio flares appeared to be correlated with the local
minima in both the \mbox{X-ray} integrated RMS variability
and the \mbox{X-ray} hardness ratio  at low energies \citep{fender09}.

The following parameters of the system were inferred from 
multiwavelength observations during outburst and quiescence: 
orbital period of  $\sim$9.15 h \citep{Garnavich00,Filippenko01},
high inclination $i\ge 60^{\circ}$ \citep{Hynes02,Zurita02},
mass function $f(M) = 7.4\pm1.1 M_{\odot}$ 
 \citep{Filippenko01}, 
source distance of $\sim$6--11 kpc \citep{Hynes02,Zurita02}.

 The values of the orbital period and  mass function were recently updated
to $6.58 \pm 0.05$ h and $f(M) = 4.5\pm 0.6 M_{\odot}$, respectively, by \cite{cs11}, based on
photometric and spectroscopic observations of the secondary star.
The source distance was instead updated to $d=4.2\pm0.5$ kpc by \citet[][hereafter ST09]{st09}  using \xte\ observations, and applying 
the scaling method to the $\Gamma$/QPO or $\Gamma$/disk normalization
diagrams. With the same procedure, these authors also determined the BH mass to be 
$M_{\rm bh}= 7.7\pm 1.3~M_{\odot}$. 
Finally, the measured optical extinction
$E(B-V)=0.58\pm0.07$
 \citep{Hynes02} is consistent with the average N$_H$ value of 
2.21$\times$10$^{21}$ cm$^{-2}$ derived from
radio maps \citep{Dickey90}.

\section{Observations and data analysis}
\label{sax_xte_analysis}

Six Target of Opportunity (TOO) observations of \j1859\  
were carried out 
from October 15, 1999, 
to March 27, 2000,  
with the \sax/Narrow-Field Instruments (NFIs). They consist of
a Low Energy Concentrator Spectrometer 
(LECS, 0.1--10 keV, Parmar et al. 1997 \nocite{Parmar97}), 
two Medium Energy Concentrator
Spectrometer (MECS, 1.3--10 keV, Boella et al. 1997), 
a High Pressure
Gas Scintillator Proportional Counter (HPGPC, 3--120 keV, 
Manzo et al. 1997)
and a Phoswich Detection System (PDS, 15-300 keV, 
Frontera et al. 1997).
During the first two TOOs the LECS was switched
off. 
The reduction of the LECS, MECS and HPGSPC data were performed with the
standard FTOOLS/SAXDAS procedures 
while data from the PDS were reduced with the 
XAS package \citep{Chiappetti97}.
The energy bands used for spectral fitting were limited
to those where the response functions are well known, i.e., 0.15--4 keV,
1.5--10 keV, 8--30 keV and 15--200 keV, for the LECS, MECS, HPGSPC and PDS,
respectively. In Table 1 we report the epoch of each TOO together with the exposure time of 
each instrument and the average count rates in the above mentioned energy bands.
A systematic error of 1\% was added in quadrature
to the statistical uncertainties of the spectral data, on the basis of the
calibration results obtained with the Crab Nebula, that was observed
on September 25, 1999, and on April 10, 2000, i.e. 20 days before the first TOO and
15 days after the last TOO.
We used the standard response matrices officially delivered for each
of the four NFIs, except for the LECS data of TOO3, TOO4, and TOO5, as where the source count rate is higher than about 50 cts s$^{-1}$,
 specific response files must be created using  
the LEMAT\footnote{http://www.asdc.asi.it/bepposax/software/saxdas/lemat.html} routine.

We accumulated total energy spectra for each TOO and for each NFI.
This procedure was justified by the fact that the spectral analysis
  showed no systematic residuals between the data
and the adopted models, meaning that the time scale of
significant source spectral evolution was longer
than the instrument integration time.
Normalization constants, allowed to vary in the recommended ranges, were introduced to take into account known differences 
in the absolute cross--calibration 
between the detectors.\footnote{ftp://ftp.asdc.asi.it/pub/sax/doc/software-docs/saxabc-v1.2.ps.gz}

A total of 131 \xte\ observations from the public data archive were analyzed,
from 1999 October 11 (MJD 51462) to 2000 July 24  (MJD 51749).
The  energy spectra and power spectral density  (PSD) were extracted 
using the standard "{\it RXTE Cookbook}" recipes in HEAsoft 6.10 analysis
software release. All \xte\ data products were
corrected for PCA deadtime. The background for energy spectra
was produced by {\it pcabackest} tool version 3.8. The PCA response
files were created using the latest release of {\it pcarmf} version 11.7
and energy-to-channel conversion table version {\it e05v04} 
\footnote{http://www.universe.nasa.gov/xrays/programs/rxte/pca/\\doc/rmf/pcarmf-11.7}. 
The PSD was computed in the 0.01-64 Hz frequency range and normalized to give a RMS deviation per Hertz. 
In the fitting procedure, we ignored PCA channels corresponding to energies below 3 
keV and above 50 keV, while the  HEXTE spectra were constrained in 20-250 keV range.
 A free  cross-calibration factor was used between PCA and 
HEXTE normalizations  resulting in an expected value of $\sim 0.85$  for both HEXTE clusters. 
A systematic error of 0.5\% was injected in the fitting process.

We also checked how many single \xte\ pointings occurred during each \sax\ TOO
and found the following results: two for TOO2 (40124-01-22-00G and 40124-01-15-00), one for TOO4 (40124-01-44-00)
and three for TOO5 (40124-01-51-00, 40122-01-04-00 and 40122-01-04-01, respectively).
We subsequently tried to produce strictly simultaneous \sax\ spectra in the
time interval corresponding to the each single \xte\ IDs.
However, because of Earth occultation time, the effective exposure time
for each NFI in these sub-intervals was too low, and eventually it
was not possible to extract spectra with a statistics good enough
to perform an accurate joint \sax/\xte\ analysis.

The spectral analysis was performed using the XSPEC package v. 12.5 \citep{arnaud96}, and
all the quoted errors for the spectral  parameters correspond to $90\%$ 
confidence level for a single parameter ($\Delta\chi^{2}=2.71$).
When fitting the \sax\ spectra, alone or joined with \xte, the interstellar absorption along the source
direction was modelled using the elemental abundance by Anders $\&$ Ebihara (1982) 
and opacities from  Morrison \& McCammon (1983).
 On the contrary, because of the lack of data below 3 keV in the PCA, the spectral
analysis of the \xte\ spectra was performed by fixing
the interstellar absorption 
to the galactic value $\nh=0.2\times 10^{22}$ cm$^{-2}$ along the source direction \citep{Dickey90}.

\begin{table*}

\caption{Best-fitting parameters of the model {\sc wabs*(dbb+bmc*highecut)} for the six \sax\ observations of \j1859.
The projected inner disk radius and luminosity are computed assuming a distance of 4.2 kpc.}

\smallskip
\begin{center}
\begin{tabular}{l|c c c c c c}
\hline
\hline
Parameter  &  TOO1  & TOO2  &  TOO3  & TOO4  &  TOO5  &  TOO6  \\
\hline
$N_{\rm H}$  ($\times 10^{22}$ cm$^{-2}$) &    0.31$^{+  0.13}_{-0.13}$       &   0.32$^{+  0.13}_{-  0.13}$  &   0.28$^{+  0.01}_{-  0.01}$   & 0.30$^{+  0.01}_{-  0.01}$  &  0.27$^{+  0.01}_{-  0.01}$ & 
 0.39$^{+  0.09}_{-  0.08}$ \\
$kT_{\rm in}$ (keV)     &  0.61$^{+  0.03}_{-  0.03}$  & 0.62$^{+  0.03}_{-  0.03}$ &  0.77$^{+  0.03}_{-  0.03}$ & 0.71$^{+  0.03}_{-  0.03}$ &
  0.59$^{+  0.03}_{-  0.03}$ &  0.16$^{+  0.03}_{-  0.02}$\\
$R_{\rm in}$ $\sqrt{{\rm cos}~i}$~ (km)               & 35$^{+3}_{-3}$  &  32$^{+3}_{-3}$ & 24$^{+2}_{-2}$ &  24$^{+2}_{-2}$ &  28$^{+3}_{-3}$  & 95$^{+84}_{- 33}$  \\
$kT_{\rm bb}$ (keV)              & 1.00$^{+  0.04}_{-  0.04}$ & 1.01$^{+  0.04}_{-  0.04}$ &  1.08$^{+  0.06}_{-  0.05}$ & 0.89$^{+  0.04}_{-  0.03}$ 
& 0.77$^{+  0.04}_{-  0.03}$ &  0.32$^{+  0.04}_{-  0.04}$\\
$\Gamma$                           &  2.15$^{+  0.04}_{-  0.04}$  &  2.15$^{+  0.04}_{-  0.04}$ & 2.17$^{+  0.08}_{-  0.09}$ & 1.80$^{+  0.07}_{-  0.08}$ & 
1.81$^{+  0.12}_{-  0.13}$  &  1.87$^{+  0.05}_{-  0.05}$\\
log(A)                          &   0.20$^{+  0.04}_{-  0.03}$   & 0.17$^{+  0.04}_{-  0.03}$ & -0.13$^{+  0.06}_{-  0.05}$&  -0.75$^{+  0.09}_{-  0.07}$ &   
-1.04$^{+  0.09}_{-  0.06}$& 0.19$^{+  0.14}_{-  0.13}$\\
$N_{\rm bmc}$            & 0.120$^{+  0.004}_{-  0.004}$  & 0.10$^{+  0.004}_{-  0.004}$ & 0.11$^{+  0.01}_{-  0.01}$ &  0.047$^{+  0.008}_{-  0.008}$ 
& 0.038$^{+  0.007}_{-  0.008}$ & 1.06$^{+  0.16}_{-  0.11} \times 10^{-3}$\\
$E_{\rm f}$ (keV)       & 77$^{+7}_{-  6}$  & 94$^{+ 11}_{-9}$ &  130$^{+ 57}_{- 32}$&  159$^{+ 72}_{- 39}$ &  $> 130$ &  --\\
$F_{\rm disk}/F_{\rm tot}$  &  0.56 & 0.56 & 0.64 & 0.77  & 0.75 & 0.67 \\
$F^a_{\rm 0.1-200~keV}$  & 3.67 & 3.23 & 3.76 & 2.37 & 1.59  & 0.10 \\
$L^b_{\rm 0.1-200~keV}$  & 7.81 & 6.90 & 8.02 & 5.05 & 3.40  & 0.21 \\
$\chi^{2}$/dof   & 131/109            & 99/105 & 133/158 & 134/150 & 160/146 & 162/138  \\
\hline
\hline
\multicolumn{7}{l}{$^a$In units of $10^{-8}$ erg cm$^{-2}$ s$^{-1}$ }\\ 
\multicolumn{7}{l}{$^b$In units of $10^{37}$ erg s$^{-1}$ } 
\end{tabular}
\end{center} 
\label{fit_dbb}
\end{table*}

\begin{figure}
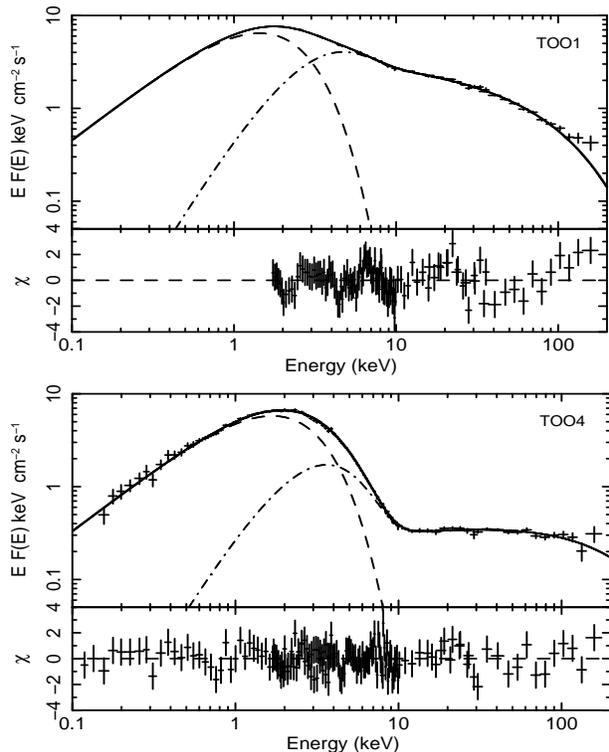

\includegraphics[ height=8cm, width=5cm, angle=-90,]{fig2.eps}
\vspace{-0.25cm}
\includegraphics[ height=8.05cm, width=5cm, angle=-90,]{fig3.eps}
\caption{Absorption-corrected deconvolved spectra in EF(E) units, best-fitting model \dbb+\bmc, and residuals between the data and the model
in units of $\sigma$ for \sax\ TOO1 and TOO4 of \j1859. The single spectral components are also plotted:
{\sc dbb} ({\it dotted line}) and {\sc bmc} ({\it dotted-dasjed line}). The emission line, marginally
evident in the first and second \sax\ 	TOO, has not been included here.}
\label{eeuf_spectra}
\end{figure}

\section{Results}
\subsection{Spectral modeling}
\label{spectral_analysis}

We modelled the source \mbox{X-ray} spectrum  with a photoelectrically-absorbed two-component model, consisting of a soft and hard feature.
For the soft component, we tested  both a simple blackbody (BB), and the multicolour-BB 
model \dbb\ in XSPEC \citep{mitsuda84}.

The latter is obtained from the convolution of several BBs at different temperatures with 
$T_{\rm bb} \propto R^{-3/4}$, where $R$ is the radial distance from the central object \citep{ss73}.
It is worth noting that the inner disk temperature ($\ktin$) and projected inner radius ($\rin$)
of \dbb\  have to be considered as order of magnitude, rather than exact values, because of
the intrinsic simplifications of the model which do not include colour-correction
factors or physical prescriptions for the inner boundary condition \citep[e.g.,][]{gierlinski08}.
Despite its simplicity, the \dbb\ model proves to be good in fitting the soft \mbox{X-ray} spectra
of both NS and BH  binary systems \citep{mcr06, lrh07}.

The high-energy component of the source \mbox{X-ray} spectrum was instead fitted with the
\bmc\ model  whose emerging spectral shape is given by
\begin{equation}
 F(E)= \frac{C_{\rm n}}{A+1} [S(E)+ A~ S(E) \ast G(E,E_0)],
\label{bmc_form}
\end{equation}

where $G(E,E_0)$ is the Green's function (GF) of the Comptonization energy operator 
\citep{tmk97}, and S(E) is the input BB spectrum. The quantity $A/(A+1)$ in equation (\ref{bmc_form}) empirically describes
the  Comptonization fraction of the input seed photons.

The free parameters of \bmc\ are the seed BB temperature $\ktbb$, the energy index $\alpha$
of the Comptonization GF, the Comptonization fraction log(A), and the
normalization.
The latter quantity, is defined as $C_{\rm n}=L_{39}/D^2_{10}$, where $L_{39}$ is the seed photon
source luminosity in units of $10^{39}$ ergs s$^{-1}$, and $D_{10}$ is the distance
in units of 10 kpc, respectively.

The GF of the \bmc\  model only depends on the spectral index $\alpha$, 
namely $G(E,E_0) \propto E^{\alpha+3}$ for $E < E_0$, and $G(E,E_0) \propto E^{-\alpha}$ for $E > E_0$.
As the GF is a broken-PL with no cut-off, the output spectrum at high energies
is yet a PL. When the data show the presence of a high-energy rollover, this must be taken into
account by multiplying \bmc\ by an e-folding factor $\propto e^{-E/E_{\rm f}}$ ({\sc highecut}).

In this sense, \bmc\ is the  most general available Comptonization model: it just measures the
relative contribution of the direct and Comptonized component and the slope
of the Comptonization spectrum, no matter which is the underlying physical process and thus 
 having no limitation in its applicability. The origin of 
the observed features is demanded in a second step to a scientific discussion.

\begin{figure}
\centerline{\includegraphics[width=9cm, height=7cm]{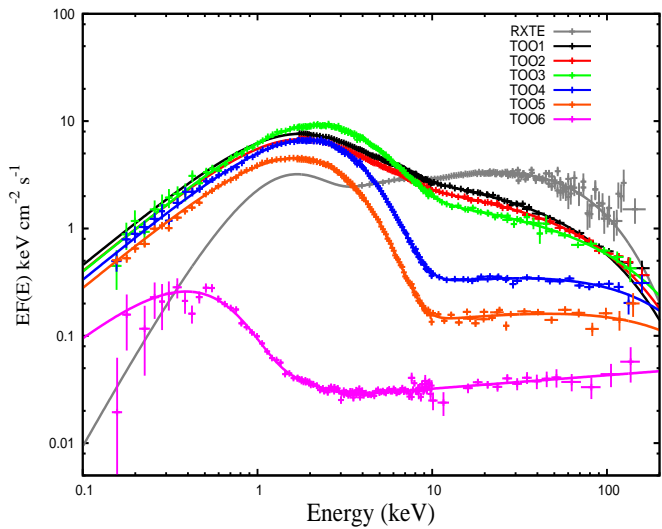}}
\caption{Spectral evolution of XTE~J1859+22 across the whole outburst. The spectra of the
six \sax\ TOOs and that corresponding to the first RXTE observation (MJD 51462) are reported.}
\label{eeuf_all}
\end{figure}

\subsection{\sax\ and \xte\ spectra}
\label{results}

The spectral analysis was first carried out with \sax, because of its  broader energy band (0.1-200 keV)
 with respect to \xte\ (3-200 keV), particularly in the critical region where interstellar absorption is present.

The \sax\ spectra of the source show the presence of a thermal component
at low energies (Fig. \ref{eeuf_spectra}). However a simple blackbody (BB) model does not provide in general
acceptable fits, while the \dbb\ better describes the data.
Thus we adopted {\sc wabs*(dbb+bmc)} as basic spectral model for the continuum emission. 

During TOO1, the presence of a high-energy cut-off in the spectrum was clearly evident. 
Indeed, the basic model  does not provide acceptable fit (\chiq/dof= 690/110), while multiplying 
\bmc\ by an e-folding factor, the model  {\sc wabs*(dbb+bmc*highecut)} gives a satisfactory result
(\chiq/dof= 131/109).

During TOO2, the spectral properties of the source were very similar to  TOO1 (Fig. \ref{eeuf_all}).  
In particular, the model {\sc wabs*(dbb+bmc)} without cut-off gives \chiq/dof=451/160, while 
with cut-off we find \chiq/dof=99/105. 

The spectrum of the source during TOO3 at higher energies was slightly harder than that of TOO1 
and TOO2. In particular,  the model {\sc wabs*(dbb+bmc)} provides \chiq/dof=167/158, while  {\sc wabs*(dbb+bmc*highecut)} 
gives \chiq/dof=133/158. The PCI for the inclusion of the cut-off components is $\sim$ 0.08, and
it was obtained through the F-test routine for discriminating among
two different spectral models for the case of multiplicative component \citep{press92}.

During TOO4, the presence  of the high-energy cut-off also  proved to be marginal (Figs. \ref{eeuf_spectra} and \ref{eeuf_all}).
The models  {\sc wabs*(dbb+bmc)} and  {\sc wabs*(dbb+bmc*highecut)} indeed give \chiq/dof=166/152 and \chiq/dof=135/151,
respectively, with PCI=0.2 in both cases.

The gradual shift towards higher energies of the spectral cut-off became evident at the time of TOO5, 
with the model {\sc wabs*(dbb+bmc)}  yielding \chiq/dof=164/147, and model {\sc wabs*(dbb+bmc*highecut)} 
 yielding \chiq/dof=160/146.
Actually, it was only possible to put a lower limit on $\ef \ga 130$ keV.

At the time of the last \sax\ observation, TOO6, the source was moving towards quiescence with a significant drop
in the bolometric luminosity. At low energies, the statistics of the source was significantly 
worse with respect to the previous observations (Fig. \ref{eeuf_all}).
At high energies, the spectrum was dominated by a pure PL component and, unlike the
case of TOO4 and TOO5, we could not put even a lower limit on the value of the cut-off 
energy $E_c$. Thus,  we just used  the model  {\sc wabs*(dbb+bmc)}
which yields \chiq/dof=162/133.

During the first two TOOs, we found some evidence for the
marginal presence of a Fe emission line (see Fig. \ref{eeuf_spectra}).
Modelling the feature with a Gaussian component added to the continuum, the fit improves to 
\chiq/dof=111/107 and \chiq/dof=85/103 for TOO1 and TOO2, respectively.
In both cases, the line width $\sigma$ was fixed to 
0.2 keV, as leaving it free during the fitting procedure, the centroid of the line shifts towards 
 too low values ($E_c \sim 6$ keV). 
Having in mind the issues reported by \cite{protassov02} about the use of F-test in the case of emission
lines, we empirically tested the significance of the feature. 
Specifically, we generated $10^4$ fake spectra for all the NFIs using for the continuum the best-fitting model
\wabs(\dbb+\bmc*{\sc highecut}) without the Gaussian emission line, and with the same time exposure of the true spectra
for each NFI.

We then fitted the simulated spectra first with the continuum model alone, and later 
with inclusion of the line (with $\sigma=0.2$ keV).
For each simulated spectrum, we computed the value $F_{12}=(\chi^2_{1} -\chi^2_{2}) /(\nu_1 - \nu_2) \nu_2/\chi^2_{2}$,
where $F_{12}$ is the probability distribution function \citep{bevington03}.
Then, we computed the number of spectra $N$ satisfying
 $F_{12} > F^{\rm obs}_{12}$, where $F^{\rm obs}_{12}$ was computed on the true data.
We found that the fraction of simulated spectra obeying the 
above condition is $38 \times 10^{-4}$ for TOO1 and $140\times 10^{-4}$ for TOO2, respectively.

These values represent the empirically-estimated  probability of chance improvement (PCI)
for the addition of the emission line.

As far as the \xte\ spectra are concerned, we note that  ST09 considered only the PCA data
in their spectral analysis of the early part of the outburst of \j1859.
The lack of spectral coverage below 3 keV and the lower energy
resolution of the PCA with respect to the LECS and MECS experiments onboard
\sax, allowed ST09 to fit the source spectrum  with a simple photoelectrically-absorbed {\sc gaussian}+{\sc bmc} model. 
However, the \sax\ low-energy data clearly
showed the presence of a soft thermal component, which
actually  carries-out  a significant part of the source luminosity (see Table \ref{fit_dbb}).
Thus, driven by the \sax\ results, we tested both \wabs(\bb\ + \bmc) and \wabs(\dbb\ + \bmc) models
also on the \xte\ spectra. We found that in both cases, they satisfactorily fit the data, 
strongly reducing the significance of the Gaussian component.
For consistency with the \sax\ results, we thus have chosen as reference
model \wabs(\dbb\ + \bmc), fixing the photoelectric absorption to the value across
the source direction $\nh=0.2 \times 10^{22}$ cm$^{-2}$.
However, given that more than one hundreds \xte\ observations were performed during the source
outburst, instead of reporting all the best-fitting parameters in a table,
we decided to plot them as a function of the MJD overplotted with
the \sax\ results (Fig. \ref{bmc_params}).

\section{Discussion}
\label{discussion}

\subsection{Spectral evolution across the outburst}

The extended temporal coverage of \j1859\ by \xte\ together with the  \sax\ observations allowed us to
deeply investigate the spectral evolution of the source during the whole outburst (Fig. \ref{eeuf_all}).

The \mbox{X-ray} continuum is characterized by the presence of a soft thermal component, described by a 
multicolour BB model (\dbb) and, at higher energies, by the convolution of a seed BB spectrum with a
Compton up-scattering GF (\bmc). 
Even if the data did not allow to fix equal BB
and \dbb\ temperatures ($\ktbb$  and $\ktin$, respectively) in the fit, we interpret them as having a common origin in the accretion disk. 
In particular, as $\ktbb > \ktin$, it is worth attributing the BB temperature  to the
innermost part of the accretion disk.
We refer to this region, where the gas gets hotter and most of Comptonization takes places,
as the transition layer or Compton cloud \citep{tf04}.
A similar spectral modelling using \bmc\  was also 
adopted by  \citet[][hereafter M09]{m09} in their \sax\ analysis of XTE~J1650-500,
even if in that case a simple BB could adequately fit the soft excess.

\begin{figure*}
\centerline{\includegraphics[height=18cm, width=15cm, angle=-90]{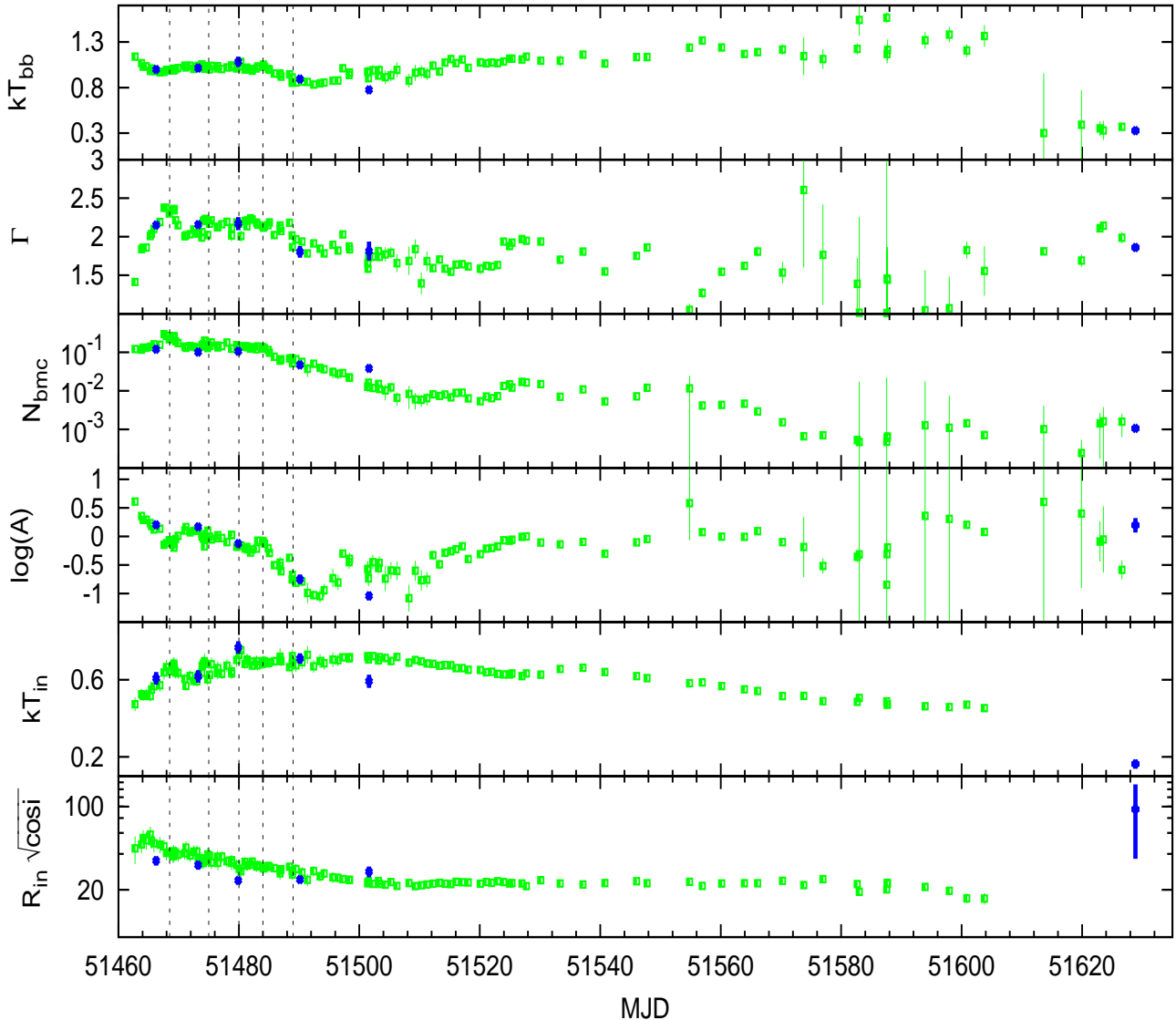}}
\caption{Evolution of the parameters of \bmc\ and \dbb\ models over the outburst of \j1859, 
for the \xte\ ({\it green points}) and  \sax\ ({\it blue points}) observations.
 The vertical dashed lines indicate the times of the five radio flares identified by B02
(see also Fig. \ref{radio_curve})}.
\label{bmc_params}
\end{figure*}

In Fig. \ref{bmc_params} we report the behaviour  of the
spectral parameters of the best-fitting model {\sc wabs*(dbb+bmc*highecut)} as a function of time (MJD),
while in Fig. \ref{radio_curve} we show the integrated (0.1-64 Hz) RMS
and the soft (2-20 keV) and hard (20-200 keV) deconvolved \mbox{X-ray} fluxes obtained in the period  where also
radio observations were available (see Section \ref{radio}).

As already reported by B02, \j1859\ entered its outburst in a hard state, with the peak of the 
soft \mbox{X-ray} light curve delayed by a few days
(Fig. \ref{radio_curve}, see also Fig. 1 in B02).
At the peak of the hard \mbox{X-ray} luminosity (MJD 51462) the source spectrum
 was characterized by a dominating thermal Comptonization (TC) feature, with cut-off energy below 100 keV 
(Fig. \ref{eeuf_all}) and photon index $\Gamma \sim 1.5$.  
In this phase the disk contribution to the total luminosity was less than 20\%, and the PSD 
showed high temporal variability (RMS $\sim$ 30\%, see Fig. \ref{radio_curve}), 
both features typical of  a canonical hard state \citep{rm06}.

During the rising phase of the soft \mbox{X-ray} flux (Fig. \ref{asm_lc}), the Comptonization spectral index $\Gamma$ 
progressively increased, reaching its maximum ($\sim 2.5$) in coincidence with the soft \mbox{X-ray} peak.
In the same period, the Comptonization fraction log(A) decreased,  as well
as the \bmc\ seed photon temperature. On the other hand an increase of the inner
disk temperature $\ktin$ occurred, together with a decrease of the projected
inner disk radius.
The simultaneous and monotonic decrease of the RMS is also  indicative of
a spectral transition from  hard to soft state.

This trend of the spectral parameters is usually expected in the scenario
of the truncated disk model \citep{done07}.

The subsequently observed flaring phase of the ASM light curve is claimed to mainly
originate in the accretion disk, which dominates the spectral emission below
10 keV (Fig. \ref{eeuf_spectra}).  
Indeed between MJD 51466 and MJD 51480, fluctuations in the inner disk temperature
(however not in its normalization) occur, and they are approximatively coincident with the secondary peaks
of the ASM light curve. 

Not surprisingly, any variation of the disk properties is expected to have physical
consequences in the radiative processes of the Comptonization region: hardening
and softening of the high-energy \mbox{X-ray} spectrum are visible in Fig. \ref{bmc_params}
through fluctuations of $\Gamma$. Once again, changes
of the geometrical extension of the corona, in response to this effect, are
still visible looking at the behaviour of log(A).
Yet, during the flaring phase, significantly less visible changes are observed in the 
$\ktbb$ value or \bmc\ normalization.

It is of particular interest  the source behaviour in the period between \sax\ TOO4 
(MJD 51490) and TOO5 (MJD 51501), 
during the smooth ASM exponential decay phase. In this phase,
  the disk contributed  more than 70\%  to the total luminosity (see Figs. \ref{eeuf_spectra}  
and \ref{eeuf_all}), and, at the same time, RMS was $\la$ 5\%, typical of a soft state.
 
However,  the photon index was $\Gamma \sim 1.8$, which is actually a value closer to
a hard state (see, e.g., XTE J1650-500 in M09). 

Note also that this time period also corresponds to the minimum value achieved by the Comptonization fraction 
parameter log(A), as shown in Fig. \ref{bmc_params},  thus  presumably in this phase 
the Compton cloud reaches its minimum size. 
As also shown by the BATSE light curve reported in B02 and by \sax\ and \xte\  in Fig. \ref{radio_curve},
 while moving towards quiescence \j1859\ experienced an increase of the hard \mbox{X-ray} flux ($> 20$ keV) around MJD 51520, 
with a corresponding increase of the  RMS ($\sim$ 10\%).

This feature however does not necessarily implies a temporary soft to hard transition at all. 
Actually, Fig.~\ref{bmc_params} shows that the rise of the hard \mbox{X-ray} luminosity 
was accompanied by an increase of the Comptonization fraction log(A) and 
a steepening of the photon index $\Gamma$. At high energies, the spectrum thus became slightly softer
but globally brighter.
This trend appears puzzling when looking at the behaviour of the two soft
seed photon components. Indeed, a slight increase of the disk temperature $\ktin$ occured a few days later (around MJD 51530)
after the new rise of the hard \mbox{X-ray} luminosity, and appears not tightly correlated with changes
in the \bmc\ parameters log(A) and $\Gamma$.

The only appreciable correlated variation is found in the \bmc\ normalization (but not in 
$\ktbb$), which actually also increased around MJD 51520 more or less
tracing the behaviour of $\Gamma$.
Thus, we interpret the rebrightening of the hard \mbox{X-ray} luminosity as an enhancement of the amount of photon
supply from the innermost region (\bmc\ normalization) such that this effect compensated,
in terms of Comptonization energy flux, the steepening of the the photon index $\Gamma$.

Note that this is a different behaviour with respect to what happened at the beginning of the outburst, 
where the coronal behaviour appeared mostly dictated by changes in the  \dbb\ component (see above).


The increased time intervals between each \xte\ pointing for MJD $\ga$ 51540 and the reduced
statistical quality of the high-energy \xte\ spectra, did not allow a thorough mapping 
of the evolution of the source spectral parameters.

However, more than five months after the outburst onset (MJD 51629), \sax\ observed the source when the 0.1-200 keV flux was $\sim 10^{-9}$ erg
cm$^{-2}$ s$^{-1}$. The broad-band coverage and sensitivity of \sax\ allowed the unique 
opportunity to catch the source
close to quiescence with an excellent data quality (Fig. \ref{eeuf_all}).
In the phase down to quiescence, the spectrum of \j1859\ is characterized by a strong PL component 
with $\Gamma \sim 1.9$ and no evidence of cut-off. The accretion disk contributes yet to a significant 
fraction of the total luminosity (about 67\%), but now is much cooler 
($\ktbb \sim 0.3$ keV, $\ktin \sim 0.2$ keV), and the Comptonization fraction is at the same level 
of the rising phase of the outburst.

The very low  temperature of the direct disk component does not allow  its detection by the PCA, and 
this explains the gap of  data points in the two lower panels of Fig. \ref{bmc_params}.
The spectral shape of the last \sax\ observation can be interpreted as a new significant 
increase of the Compton cloud characteristic size, together with a very high electron temperature, 
 consequences of a strongly reduced efficiency in Compton cooling by the seed photons. 

\subsection{Justification of the adopted spectral model for describing the high-energy emission}

The high-energy emission of BH sources in different spectral states has been fitted with several 
 XSPEC Comptonization models such as \comptt\ \citep{cadolle06}, \compps\
\citep{frontera01}, or \eqpair\ \citep{delsanto08}.
All the above mentioned models contain a detailed treatment of the radiative transfer physics,
but each of them present its own drawbacks.

In particular, it is worth emphasizing that both \comptt\ \citep{t94} and \compps\ \citep{ps96}
are \emph{static thermal Comptonization models} with different ranges of applicability for
the plasma electron temperature and optical depth (preferably low $\kte$ and high $\tau$ for
\comptt\ and the opposite for \compps).
On the other hand \eqpair\ \citep{coppi99} allows to compute Comptonization spectra for
a plasma which can be purely thermal or hybrid (thermal plus non-thermal electron distribution).

Thus, when one of such models is used to fit the source spectra
at high energies, it is implicitly assumed that the spectral changes occur
in the framework of static-only thermal Comptonization processes (\comptt, \compps)
or due to a switch from static thermal to hybrid Comptonization (\eqpair).

By fitting the \xte\ spectrum of \j1859\ during the bright hard state at the
beginning of the outburst (MJD 51462, see Fig. \ref{eeuf_all}) with \dbb+\compps\ model, 
assuming spherical geometry of the Compton cloud,  we found $\kte \sim 30$ keV and $\tau \sim 1$
(with Comptoniztion parameter $Y \approx 0.2$).
Following the source evolution from the hard to the soft state we note however
that, e.g.,  the fit of the \sax\ spectra provided an electron temperature $\kte$ increasing from $\sim$ 50 keV (TOO1) to $\sim$ 200 keV  (TOO5) with a corresponding decrease of the
optical depth from $\tau \sim$ 1 to $\sim$ 0.2, respectively, and  $Y \approx 0.3-0.4$.

The $\chi^2$ values were acceptable in all cases, but in our opinion it is
difficult to explain such a behaviour during a hard to soft transition.
In particular, the enhanced disk contribution to the total
luminosity (see Table \ref{fit_dbb}) is expected to provide strong Compton cooling of the coronal electrons 
and their temperature should decrease rather than increase. 

It is also worth mentioning that because of the \xte\ and \sax\ high-energy threshold
around  200 keV it was not possible to find evidence of any additional PL-like
component above the energy rollover such as reported, e.g. in \mbox{GX 334--4} by
\cite{delsanto08} with INTEGRAL/SPI. Thus, hybrid Comptonization (\eqpair), rather than
excluded, is not required by the data.
The progressive increase of the high-energy cut-off during the hard to soft transition, together
with the spectral index steepening, can be more easily explained by the transition from a 
TC to bulk Comptonization (BC) dominated state (e.g., M09).

Unfortunately, specific models for BC in BH sources are not yet available. The \comptb\
model of \cite{farinelli08} is based on an assumption of sub-relativistic bulk motion,
expected to hold in NS systems due to strong radiation pressure from the NS surface,
but not in BHs. 

Actually, our choice of  \bmc\  was mostly motivated by the fact that 
with its parametrization, it represents the best compromise between a phenomenological
and physical description of the high-energy source spectral behaviour, possibly 
avoiding contradictory results which may obtained, e.g., using
static TC models.

\subsection{Correlation between X-ray and radio emission}
\label{radio}

\begin{figure}
\centerline{\includegraphics[width=9cm, height=10cm]{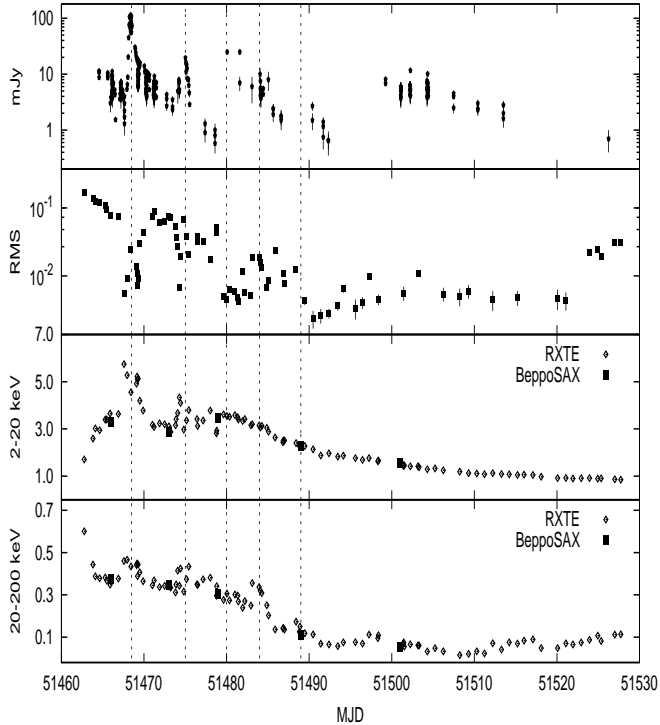}}
\caption{From top to bottom: radio light curve at all wavelengths (see Section \ref{radio}), integrated 0.1-64 Hz RMS, 2-20 keV and 20-200 keV
fluxes in units of $10^{-9}$ erg cm$^{-2}$  s$^{-1}$ obtained from the deconvolved best fit model \dbb+\bmc\
 using  both  \sax\ and \xte. The time period of the plotted data corresponds to that of the available radio observations.}
\label{radio_curve}
\end{figure}

\begin{figure}
\centerline{\includegraphics[width=9cm, height=8cm]{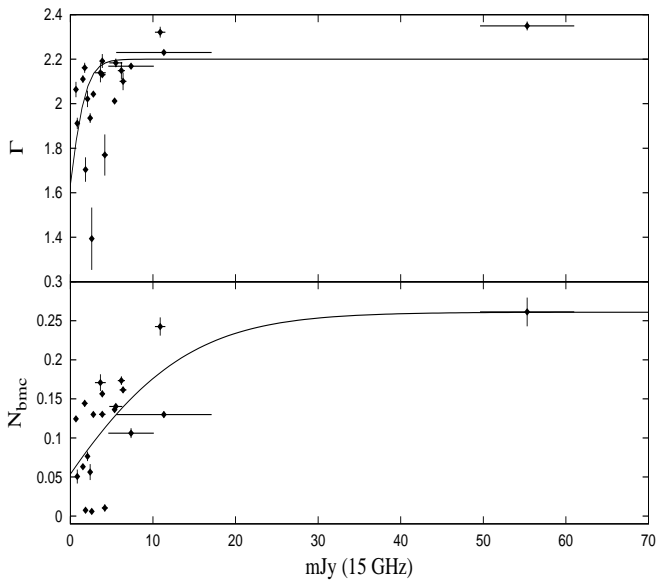}}
\caption{Spectral photon index $\Gamma$ ({\it upper panel}) and \bmc\ normalization ({\it lower panel}) as a function of the
 radio flux normalized at 15 GHz for \j1859. Data have been averaged over one day.}
\label{alphabmc_vs_radio}
\end{figure}

A thorough  study of the correlation between the \mbox{X-ray} and radio
emission of \j1859\ was previously carried-out by B02.
In Fig. \ref{radio_curve} we report the cumulative light curve of the source
obtained from several observing facilities at frequencies ranging from 1.22 GHz to 22.5 GHz
in the period from 1999 October 13 (MJD 51464) to 1999 December 13 (MJD 51525),
and we address the reader to B02  for details.

The radio light curve shows a rather complex behaviour, with fluxes ranging from 0.5 to 100 mJy,
and five major flares identified by B02 (drawn 
as vertical lines in Fig. \ref{bmc_params} and \ref{radio_curve}) occurring 
over an otherwise globally fading behaviour.

Starting from the results of the \mbox{X-ray} spectral analysis, we searched
for correlated patterns  between the radio emission and the source spectral parameters
in spite of integrated \mbox{X-ray} fluxes.

To avoid biases in the obtained results, due to different observing times at different frequencies,
we preliminary  normalized the radio fluxes to those at 15 GHz, using
the relation $F(15~{\rm GHz})=F(\nu) (15~{\rm GHz}/\nu)^{\alpha}$ mJy, with the spectral index 
$\alpha$ reported in Table
1 of B02. We then considered the normalized radio light curve and single  parameters of the \dbb+\bmc\ model
averaged over half and one day, respectively, and we calculated the Spearman correlation coefficient $r$.
Because the number of points was more than 20, we also calculated the significance of the correlation
using the associated $t$-variable of the Student distribution.

The strongest X--radio correlation ($> 99$\% confidence level)  is found, over one-day time scale, 
between the radio flux and both the spectral photon index $\Gamma$ and  \bmc\ normalization ($\nbmc$), 
with  $r \sim 0.6$ in both cases (Fig. \ref{alphabmc_vs_radio}).

On the other hand, the hard \mbox{X-ray} flux (given by the convolution term in equation [\ref{bmc_form}])
is proportional to $\nbmc$, thus in turn its 
positive correlation with the radio emission is eventually expected as well.
We find indeed  that it is similar 
to that observed with the aim of BATSE data (see Fig. 8 in B02).

It is interesting to observe that brighter and steeper \mbox{X-ray} spectra are associated
 with brighter radio events, with a possible \emph{saturation effect} of the photon index $\Gamma$
for increasing radio fluxes.

\begin{figure}
\centerline{\includegraphics[height=0.5\textwidth, width=0.4\textwidth, angle=-90]{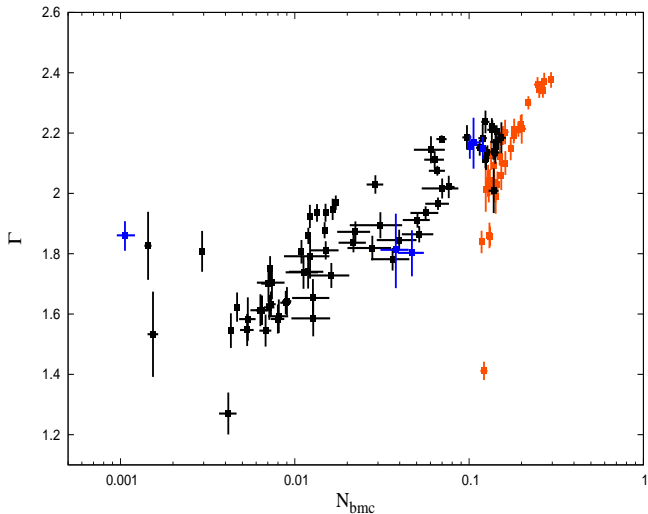}}
\caption{Spectral photon index $\Gamma$  versus \bmc\ normalization  
 for the \xte\ data set analyzed by ST09 ({\it orange}), plus the second part of the outburst 
({\it black}) and the six \sax\ TOOs (\emph{blue}). For clarity purposes we
plotted only the points for which the ratio between the best-fitting value and
 the 1$\sigma$ error determined by XSPEC is $\ga$ 3.}
\label{indexnorm}
\end{figure}

As reported in Fig. \ref{indexnorm}, the same saturation effect is instead not observed in a diagram of $\Gamma$
 vs. $\nbmc$, a typical feature of most, albeit not all, BHCs (ST09).

The index saturation in a $\Gamma$ vs. $N_{\rm bmc}$ or $\Gamma$ vs. $\nu_{\rm QPO}$ diagram
 is interpreted as an indirect evidence for the presence of a converging flow (bulk motion)
 close to the central object in BH sources, and it is expected to occur  when the optical depth $\tau \gg 1 $ 
  \citep{lt07, ts09}.

The thorough monitoring with \xte\ during the whole  outburst  of \j1859\ (Fig. \ref{bmc_params})
seems to exclude the possibility of an incomplete sampling problem. 

As $\tau \propto \dot{M}$, one possibility, albeit speculative, is that in this source the accretion rate $\dot{M}$
does not exceeds the critical value above which the Comptonization parameter $Y$, and in turn
 the spectral index $\Gamma$, becomes insensitive to $\tau$.

The apparent index saturation in the $\Gamma$ vs. radio diagram is thus challenging in this
context.

Correlation patterns between the radio emission and the integrated RMS variability were investigated
by \cite{fender09} who found that the five radio flares identified by B02 occurred around periods
 of reduced RMS. This is almost evident looking at Fig. \ref{radio_curve}, albeit the
 correlation is much less significant ($r \la 0.2$). 
Also, the different behaviour of GX 339--4 and possibly XTE J1550--564 seems
 to point against a direct casual link.


\begin{figure}
\centerline{\includegraphics[ height=0.5\textwidth, width=0.4\textwidth, angle=-90]{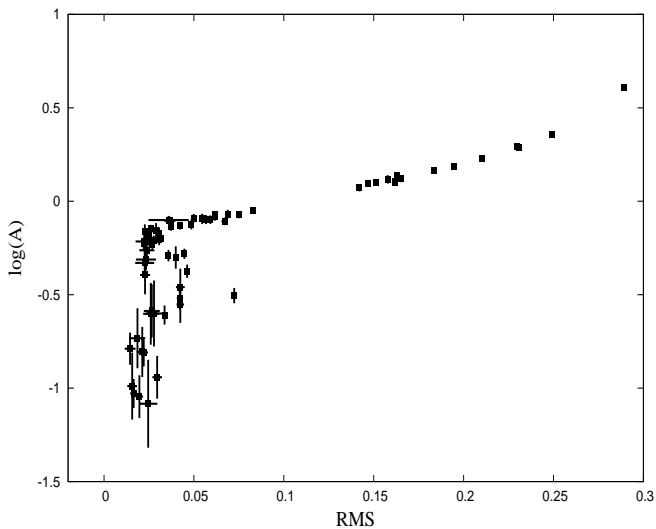}}
\caption{Fraction of Comptonized seed photons, parametrized through log(A) in \bmc, versus 
 integrated 0.1-64 Hz RMS variability using \xte\ data.}
\label{loga_vs_rms}
\end{figure}

\subsection{Transition layer as a formation site for the fast X-ray variability}

One of the most established observational results of BH systems is that their PSD  strongly
 correlates with the source spectral state. During the hard state, the RMS variability is high
(up to 40\%) and the PSD is described by a broken PL with a flat plateau up to the break frequency.
When the source moves progressively towards the intermediate and soft state, the RMS variability 
decreases below 10\% and the PSD is a PL with index 1.0-1.5 \citep{rm06}.
The rapid variability down to $\sim 10^{-2}$ s is probably mostly dictated by variations of the 
Comptonization region (TL), as the characteristic viscous time-scale of the accretion disk 
is several hundreds of seconds \citep{churazov01}.

Models for explaining the rapid variability as a function of the source spectral state have been
proposed, e.g.,  by \cite{tsa07} and \cite{ingram11} in the framework of propagating perturbations
in bounded configurations.

In order to strenghten the hypothesis that rapid variability originates in the TL, 
in Fig. \ref{loga_vs_rms} we present a plot of the log(A) parameter of \bmc\ 
versus the integrated 0.1-64 Hz RMS of \j1859. A tight  positive correlation is observed between 
the two quantities  for RMS $\ga$ 5\%.
The Spearman rank correlation coefficient in this case is $r=0.95$,  while including the whole data 
it is $r=0.83$. 
As it can be seen from the definition of the \bmc\ model (Eq. [\ref{bmc_form}]), log(A) quantifies
the  contribution of the Comptonized spectrum with respect to the input seed photon population.
To first approximation, log(A) is an indirect  measurement of the size
of the Compton cloud. 
The fact that when log(A) increases, the corresponding source RMS increases as well, is thus a strong
 additional probe that it is the Compton cloud the site of origin of the rapid variabilty.

On the other hand, unlike the case of the broad-band noise component, the origin of low-frequency
QPOs  both in NS (10--60 Hz) and BH (0.1--10 Hz) sources, is still controversials, as either models 
proposing a Lense-Thirring disk precession origin 
\citep[e.g.,][]{stella98,stella99,schnittman05, schnittman06} or magnetoacoustic
 oscillations of the Comptonization region \citep[e.g.,][]{to99, wagoner01} have proved to be in good 
agreement with observations.

A tight correlation between the spectral photon index $\Gamma$ of the Compton cloud  
and characteristic C-type QPOs, which have variable frequency (0.1--10 Hz) and high coherence parameter
($Q \ga 10$), has been observed in a large sample of BH sources by ST09.
Additionally, C04 (see their Fig. 3) found a  tight anti-correlation between the RMS and  $\nu_{\rm QPO}$
for QPOs of the same type. 
These observational features, together with the fact that usually QPOs show greater amplitude
at higher energies, apparently would favour the Compton cloud, rather than the accretion disk, as
the region where QPOs form.

However, it is worth noting that the above mentioned correlations are not in contradiction with
a tilted disk origin, if most of the disk thermal soft radiation is intercepted and Comptonized
by the hotter inner region.
This interaction indeed, would shift to higher energy the soft photons modulated at the Lense-Thirring
 disk precessional frequency, at the same time preserving the $\Gamma$ vs. $\nu_{\rm QPO}$
and RMS vs. $\nu_{\rm QPO}$ relations.

It is also worth mentioning that no correlation with the RMS was  observed by C04 for  A-type 
and B-type QPOs (7 $\la \nu \la$ 8 Hz,  $Q \la $3 and $4.5 \la \nu \la 6.5$ Hz, $Q \ga 6$, respectively)
and that the $\Gamma$ vs. QPO correlations diagrams for BH mass determination with the scaling method
were produced by ST09 using only Type-C QPOs.

This pattern behaviour could be indicative of a different physical origin for the various types of
QPOs with Type-C ones associated to magneto-acoustic oscillations of the TL \citep{tf04}
and Type-A and Type-B ones originating in formation site either 
away from the TL or in some physical process (e.g., jet/corona connection) almost independent
on the geometrical variations of the TL itself.

\section{Conclusions}
\label{conclusions}

The joint \sax/\xte\ analysis that we performed on the \mbox{X-ray} Nova \j1859 during its 1999 outburst has demonstrated
the importance of getting simultaneous spectral and temporal informations during the source evolution.
The extensive observational campaign performed by the two satellites indeed allowed us to map thoroughly \j1859\
from the onset of its outburst down a phase close to quiescence.
We modelled the hard \mbox{X-ray} component with the \bmc\ model which has the advantage
of being a fully generic Comptonization model, suitable to be applied in any
physical regime because the shape of the up-scattering
GF function in \bmc\ can be considered as general as possible. 

The source experienced all the canonical spectral states of BHCs from hard to soft 
(Fig. \ref{eeuf_all}).
During the soft state, with a disk contribution to the total 
luminosity higher than 70\% and an RMS $<$ 10\%, the high-energy spectral index, with $\Gamma \sim 1.8$, 
was actually closer to typical values of the hard state. 

A significant correlation was observed between $\Gamma$ and the radio emission, with a possible
effect of index saturation at the brightest radio emission levels.
Additionally,  we found a tight correlation between the Comptonization fraction
of the seed photons, parametrized through log(A), and the source RMS, which 
allowed us to strengthen the claim that most of the source variability originates in the transition layer, 
where also Comptonization is the dominating process.
This correlation is new and can be of key interest in understanding
the accretion geometry of \mbox{X-ray} binary systems systems hosting either a BH or a NS.

We thus strongly suggest to test it on a large sample of
sources, which is possible due to the large amount of archival
\xte\ data.

\section*{Acknowledgments}

RF is grateful to Lev Titarchuk for useful discussions related to
the spectral evolution of \mbox{X-ray novae}, and aknowledges financial support from agreement ASI-INAF  I/009/10/0.
 The authors also acknowledge the anonymous referee
whose detailed report was helpful in better outlining the most important results of the
paper.

\bibliographystyle{mn2e}
\bibliography{biblio}

\label{lastpage}

\end{document}